\documentclass[%
 reprint, superscriptaddress,
 amsmath,amssymb,
 aps,
]{revtex4-2}

\usepackage{graphicx}
\usepackage{dcolumn}
\usepackage{bm}
\usepackage[colorlinks=true,citecolor=blue,linkcolor=blue]{hyperref}

\usepackage{booktabs}

\begin{document}

\preprint{APS/123-QED}

\title{BN-embedded monolayer graphene with tunable electronic and topological properties}





\author{Chih-Piao Chuu}
\thanks{These authors contributed equally to this work.}
\affiliation{Institute of Atomic and Molecular Sciences, Academia Sinica, Taipei 10617, Taiwan}
\affiliation{Physics Division, National Center for Theoretical Sciences, Hsinchu 300, Taiwan}
\affiliation{Department of Electrophysics, National Yang Ming Chiao Tung University, Hsinchu, 30010, Taiwan}

\author{Wei-En Tseng}
\thanks{These authors contributed equally to this work.}
\affiliation{Institute of Atomic and Molecular Sciences, Academia Sinica, Taipei 10617, Taiwan}
\affiliation{Department of Physics, National Taiwan University, Taipei 10617, Taiwan}

\author{Kuan-Hung Liu}
\affiliation{Institute of Atomic and Molecular Sciences, Academia Sinica, Taipei 10617, Taiwan}
\affiliation{Department of Physics, National Taiwan University, Taipei 10617, Taiwan}

\author{Ching-Ming Wei}
\affiliation{Institute of Atomic and Molecular Sciences, Academia Sinica, Taipei 10617, Taiwan}

\author{Mei-Yin Chou}
\email[Corresponding author. Email: ]{mychou6@gate.sinica.edu.tw}
\affiliation{Institute of Atomic and Molecular Sciences, Academia Sinica, Taipei 10617, Taiwan}
\affiliation{Department of Physics, National Taiwan University, Taipei 10617, Taiwan}
\affiliation{School of Physics, Georgia Institute of Technology, Atlanta, Georgia 30327, USA}

\date{\today}

\begin{abstract}

Finding an effective and controllable way to create a sizable energy gap in graphene-based systems has been a challenging topic of intensive research. We propose that the hybrid of boron nitride and graphene ($h$-BNC) at low BN doping serves as an ideal platform for band-gap engineering and valleytronic applications. We report a systematic first-principles study of the atomic configurations and band gap opening for energetically favorable BN patches embedded in graphene. Based on first-principles calculations, we construct a tight-binding model to simulate general doping configurations in large supercells. Unexpectedly, the calculations find a linear dependence of the band gap on the effective BN concentration at low doping, arising from an induced effective on-site energy difference at the two C sublattices as they are substituted by B and N dopants alternately. The significant and tunable band gap of a few hundred meVs, with preserved topological properties of graphene and feasible sample preparation in the laboratory, presents great opportunities to realize valley physics applications in graphene systems at room temperature.

\end{abstract}


\maketitle

\section{Introduction}
For device applications of graphene, it is highly desirable to open a band gap in a controllable manner without strongly perturbing its intrinsic property. Recently, successful synthesis of monolayer graphene containing co-doped B and N has been reported by either chemical vapor deposition (CVD) or direct local chemical conversion \cite{Ci10,Chang13,Muchharla13,Telychko15,Lu13,Gong14,Ba17,hBNC1,hBNC4}. This novel two-dimensional material, a hybrid of boron nitride and graphene ($h$-BNC), is a semiconductor, distinctly different from its parent materials of gapless graphene and the large-gap hexagonal boron-nitride ($h$-BN) monolayer. A band gap of 18 meV was first reported for a $h$-BNC nanoribbon with 44$\%$ BN doping prepared by using methane and ammonia borane (NH$_3$-BH$_3$) precursors in the CVD growth \cite{Ci10}. At this high concentration, both graphene and $h$-BN domains were present and phase separated, hence the gap opening was explained by the quantum confinement effect in graphene. In contrast, in CVD samples with a lower BN concentration and more dispersed substitution, band gaps of 200 meV and 600 meV were found for 2$\%$ and 6$\%$ BN-embedded samples, respectively \cite{Chang13}.
More recently, CVD growth using different single-source precursors containing C, B, and N atoms obtained homogeneous $h$-BNC structures \cite{hBNC1,hBNC4}, and a significant band gap between 1.4 and 1.6 eV was observed at about 17$\%$ BN concentration \cite{hBNC4}. As will be discussed in this work, this BN-embedded graphene ($h$-BNC) system turns out to represent a particularly interesting hybrid system with unprecedentedly favorable electronic and topological properties.

To utilize the electronic and topological properties of the semiconducting $h$-BNC monolayer controlled by the BN concentration, a phase separation between graphene and $h$-BN needs to be avoided during the growth. It has been demonstrated that the growth of $h$-BNC can be achieved under high temperature and non-equilibrium conditions by chemical conversion methods \cite{Gong14}. In addition, it was observed experimentally that N dopants occupy primarily the same sublattice in N-doped graphene on the copper substrate \cite{Zhao13}. Therefore, N dopants could possibly function as a seed to attract B as in the direct chemical conversion process \cite{Gong14}, controlling the sublattice polarization of BN dopants throughout the sample.

Several previous calculations on the band gap opening in $h$-BNC were reported in the literature \cite{Zhao12,Shinde11,Xu10,Manna11,Fan12,Dvorak14,Zhang11,Kaloni14,Sirikumara16,Nascimento15,theory5,theory4,theory2}. Some studies had a large concentration of BN, hence with quantum dots or nanoribbons embedded in graphene,  the size of the energy gap was explained by the quantum confinement effect related to the width of carbon walls between them \cite{Zhao12,Shinde11}.  Other studies focusing on different configurations of BN patches found that the band gap values varied significantly with their size and shape \cite{Xu10,Manna11,Nascimento15,theory5}, as well as the orientation of single B-N pairs \cite{theory2}. In one case, the doping-induced gap was found to vary over an order of magnitude even for a given BN concentration \cite{Nascimento15}. Another study found an almost linear increase of the band gap with the BN concentration up to 75\% \cite{Kaloni14}. Since not all configurations used in the calculations were energetically optimized, it may not be straightforward to compare the calculated energy-gap results with experiment. In addition, no clear physical picture has been provided for the gap variation with the BN concentration.

In the present study, we report a comprehensive study of the underlying mechanism of gap opening in graphene upon BN doping. We focus on the low-concentration range where the graphene-like bands are preserved even with a finite gap, and consider only energetically favored atomic configurations. Our first-principles calculations demonstrate that the band gap in the low doping region of $h$-BNC actually varies linearly with the ``effective" BN concentration (to be defined later). Additionally, a tight-binding model is constructed to simulate general doping configurations in order to eliminate the finite-size effect in small supercells. We study in detail the linear variation of the band gap and provide the physical explanation in terms of symmetry breaking on the sublattice in graphene. It turns out that the intrinsic topological properties of graphene are in fact preserved, leading to nonzero Berry curvatures at the Dirac points when the gap is opened. This makes gapped $h$-BNC a superior system to explore valley physics in two dimensions, including the valley Hall effect (VHE) and other topological transport properties \cite{Xiao99,Lensky15,Gorbachev14}.  The linear dependence of the band gap on the effective BN concentration provides a control parameter to tune the properties of the system, similar to the role played by the gate voltage in bilayer gapped Dirac materials \cite{Sui15,Shimazaki15}. The noticeable gap size caused by the broken sublattice symmetry presents a unique platform allowing for the operation of valleytronics at room temperature.

\section{Calculational Methods}
Our first-principles calculation is based on density functional theory (DFT) with the local density approximation (LDA) using the projected argmented-wave (PAW) method \cite{Blochl94} as implemented in the Vienna Ab initio Simulation Package (VASP) \cite{Kresse96}. The electronic wave functions are expanded in plane waves with a cutoff energy of 400 eV. A unit cell containing the atomic layer and a vacuum region of 15 $\textrm{\AA}$ is used. We adopt the lattice constant of graphene in the calculation, since the BN patches are the minority component. Various in-plane supercells are used depending on the concentration of B-N pairs. The 8$\times$8$\times$1 k-point mesh is used for the 3$\times$3 supercell, 6$\times$6$\times$1 k-points for the 4$\times$4 supercell, 5$\times$5$\times$1 k-points for the 5$\times$5 supercell, 4$\times$4$\times$1 k-points for the 6$\times$6 and 7$\times$7 supercells, and 3$\times$3$\times$1 k-points for the 8$\times$8, 9$\times$9, and 10$\times$10 supercells, respectively. The energy self-consistency is at the level of $10^{-5}$ eV per supercell. For structural optimization, all atoms are relaxed until the forces reach $10^{-2}$ eV/$\textrm{\AA}$. The GW$_0$ quasiparticle band gaps \cite{Louie97} are calculated with three iterations of the quasi-particle energies in the Greens functions for a B$_3$N$_3$ ring inserted in 4$\times$4 and 5$\times$5 supercells. 
We extract the quasiparticle band gap at the infinite separation of the layers based on the procedure in quasiparticle band-gap calculations for the $h$-BN monolayer \cite{Berseneva13}. Due to the significant larger unit cell of $h$-BNC compared with that of $h$-BN, we used an energy cut-off of 300 eV and increased the number of total bands in the calculation proportionally with the vacuum thickness.

Our tight-binding (TB) model takes into consideration only the nearest-neighbor couplings, which are sufficient to capture the low-energy electronic properties of $h$-BNC. The parameters were determined by fitting to calculations using density functional theory (DFT). We used a least-square fit to the low energy bands of monolayer graphene to obtain the nearest-neighbor coupling between two carbon sites $t_{CC}=2.52 $ eV. Next, the on-site energy difference between B and N atoms, $\Delta E=E_B-E_N$ and the B-N coupling $t_{BN}$ were determined by fitting to the DFT band structure of monolayer $h$-BN. We obtained $\Delta E=4.6$ eV from the energy gap at the $K$ point and $t_{BN}=2.48 $ eV from the band dispersion.
The on-site energies $E_B$ and $E_N$ are set to be $(\Delta E)/2$ and $-(\Delta E)/2$, respectively, with $E_C=0$, which simplifies the fitting process while giving reasonable and satisfying fitting results. Finally, a least-square fit to the DFT band gaps of various $h$-BNC configurations was performed to determine the two optimized parameters given as $t_{CB}=1.85 $ eV and $t_{CN}=1.78 $ eV. The tight-binding parameters used in this study are summarized in Table \ref{table:1}. 
As shown in Appendix, the energy gaps from tight-binding (TB) calculations are in good agreement with those from DFT calculations for various $h$-BNC configurations used in the fitting.
\begin{table}[tb]
\caption{Tight-binding parameters obtained by fitting to DFT calculations. The on-site energy of C atoms $E_C$ is set to zero, and only the nearest neighbor couplings $t$ are considered.}
\label{table:1} 
\centering
\addtolength{\tabcolsep}{2.5pt}
\begin{tabular}[t]{cccccccc}
    \hline
     & $E_C$ & $E_B$ & $E_N$ & $t_{CC}$& $t_{CB}$ & $t_{CN} $& $t_{BN}$\\
    \hline
    Energy (eV) & 0.0 & 2.30 & -2.30 & 2.52 & 1.85 &1.78 &2.48\\
    \hline
\end{tabular}
\end{table}

\section{Results and Discussions}

\subsection{Energetics}
In this study, we only consider the configurations containing an equal amount of B and N. This is the situation when the CVD synthesis uses ammonia borane or borazine. Before investigating the electronic properties, we first determine the energetically favorable atomic configurations at each doping level with the BN concentration restricted to less than 20$\%$. We find that the nearest-neighbor B-N pair configuration is highly favorable energetically, with an energy at least 1 eV lower than that of well-separated B and N atoms. It is therefore reasonable to consider the B-N pair as the building block when constructing BN nano-patches from the energetics perspective.

\begin{figure}[ht]
\centering
\includegraphics[width=3.2in]{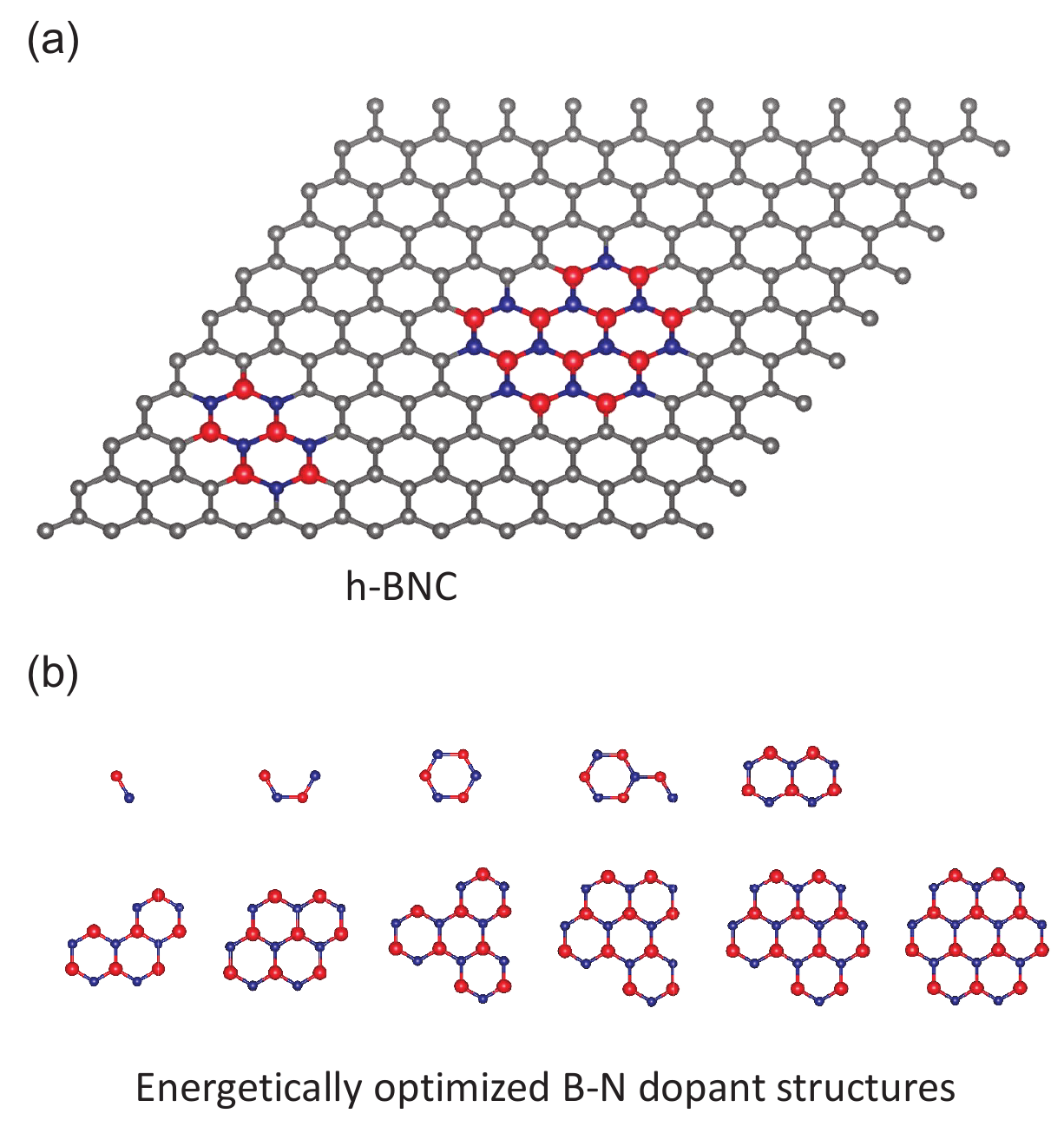}
\caption{(a) Schematic formation of an $h$-BNC monolayer. (b) Energetically optimized structures of one to 12 B-N pairs inserted into graphene. The red and blue spheres represent B and N atoms, respectively}\label{bncstruc}
\end{figure}

Figure 1 shows the energetically favorable structures for various numbers of B-N pairs calculated using different graphene supercells. The B-N pairs prefer to stick together, and the completion of hexagonal rings is mostly favorable. As the number of B-N pairs increases, the shape of the optimized structure also changes. In addition to the lowest-energy configurations shown in Fig. 1, other possible structures exist with energy differences per B-N pair smaller than the thermal energy at room temperature. To quantitatively study the energetics factors affecting the formation of BN dopants, we evaluate the formation energy $E_m$ defined as
\begin{eqnarray}
E_m&=&E-\sum_i\mu_in_i=E-\mu_{\textrm{B-N}}N_{\textrm{B-N}}-\mu_\textrm{C}N_\textrm{C}
\end{eqnarray}
for an $N_s\times N_s$ graphene supercell with $N_{\textrm{B-N}}$ B-N pairs embedded and $N_\textrm{C}=2N_s^2-2N_{\textrm{B-N}}$ carbon atoms remaining. $E$ is the total energy per supercell, and the chemical potentials $\mu_{\textrm{B-N}}$ and $\mu_\textrm{C}$ are the energy per B-N pair in an $h$-BN monolayer and the energy per C atom in graphene, respectively. The total BN concentration is then defined as $c_{BN}=N_{\textrm{B-N}}/N_s^2$.

\begin{figure}[ht]
\centering
\includegraphics[width=3.4in]{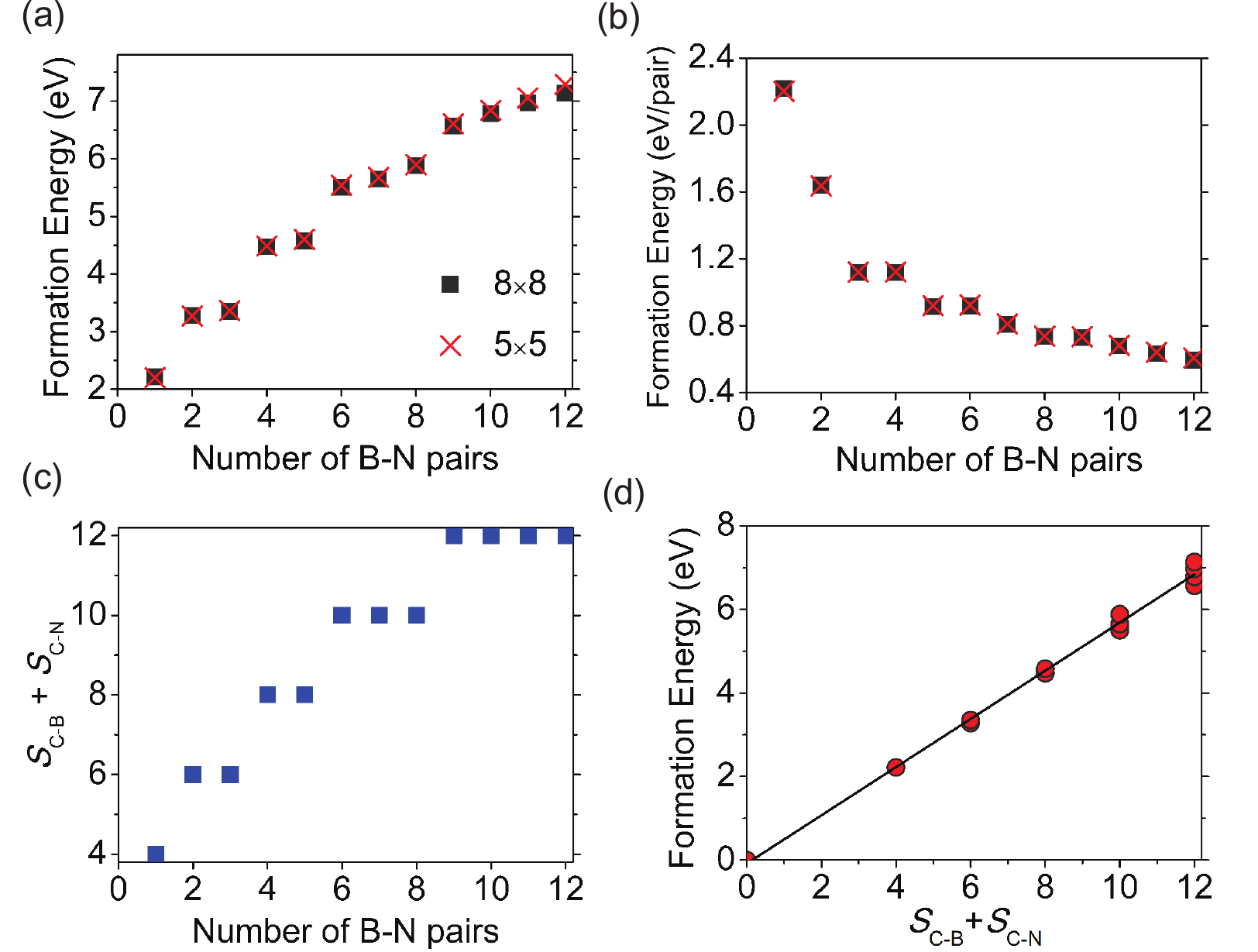}
\caption{Dependence among the formation energy, number of B-N pairs, and interfacial bonds. The formation energy (see text) is evaluated for the optimized $h$-BNC structures in Fig. 1. (a) The total formation energy and (b) the normalized formation energy per B-N pair are calculated using 5$\times$5 (triangles) and 8$\times$8 (squares) supercells. (c) The number of interfacial bonds as a function of the size of BN patches. (d) The formation energy as a function of the sum of the numbers of C-B and C-N bonds.}\label{bncformation}
\end{figure}

Figure 2(a) shows the formation energy results as a function of the number of B-N pairs for the optimized structures in Fig. 1, while Fig. 2(b) plots the average formation energy per pair. The total formation energy (as well as the normalized value per B-N pair) is primarily determined by local BN structures and nearly independent of the size of the supercells, even for large doping concentration (48$\%$ with 12 B-N pairs in a 5$\times$5 supercell). The value is positive, indicating that it costs energy to embed BN patches in graphene. The plateaus in Fig. 2(a) correspond to the noticeable drops of the average formation energy per B-N pair in Fig. 2(b). For example, a significant energy cost reduction of 0.5 eV and 0.2 eV per pair occurs when the number of B-N pairs increases by one from two to three and from four to five, respectively. This happens when the lowest-energy structure completes the formation of B-N hexagonal rings, which reduces the interface boundary in the BN patch and eliminates the net electrostatic dipoles. It turns out that the closed structure of $\text{B}_8\text{N}_8$ was observed in the CVD growth of $h$-BNC \cite{hBNC4}. 

In addition, the magnitude of the formation energy is closely correlated with the total number of interfacial bonds (both C-B and C-N bonds). Figure 2(c) plots the variation of the number of interfacial bonds $S_{\textrm{C-B}}$ + $S_{\textrm{C-N}}$ for the structures in Fig. 1. We can see a similar behavior in Figs. 2(a) and 2(c). Therefore, as shown in Fig. 2(d), the formation energy increases linearly with the number of interfacial bonds. For a fixed number of B-N pairs, the structure with the least number of C-B and C-N bonds gives the lowest formation energy, indicating that phase separation between BN patches and graphene domains is energetically favorable as observed in the CVD growth of $h$-BNC thin films \cite{Ci10,Chang13}. However, at low BN concentration and under non-equilibrium growth conditions, it is possible that BN patches are dispersed throughout the graphene layer, which is the interesting system we are focusing on.

\begin{figure*}
\centering
\includegraphics[width=.59\linewidth]{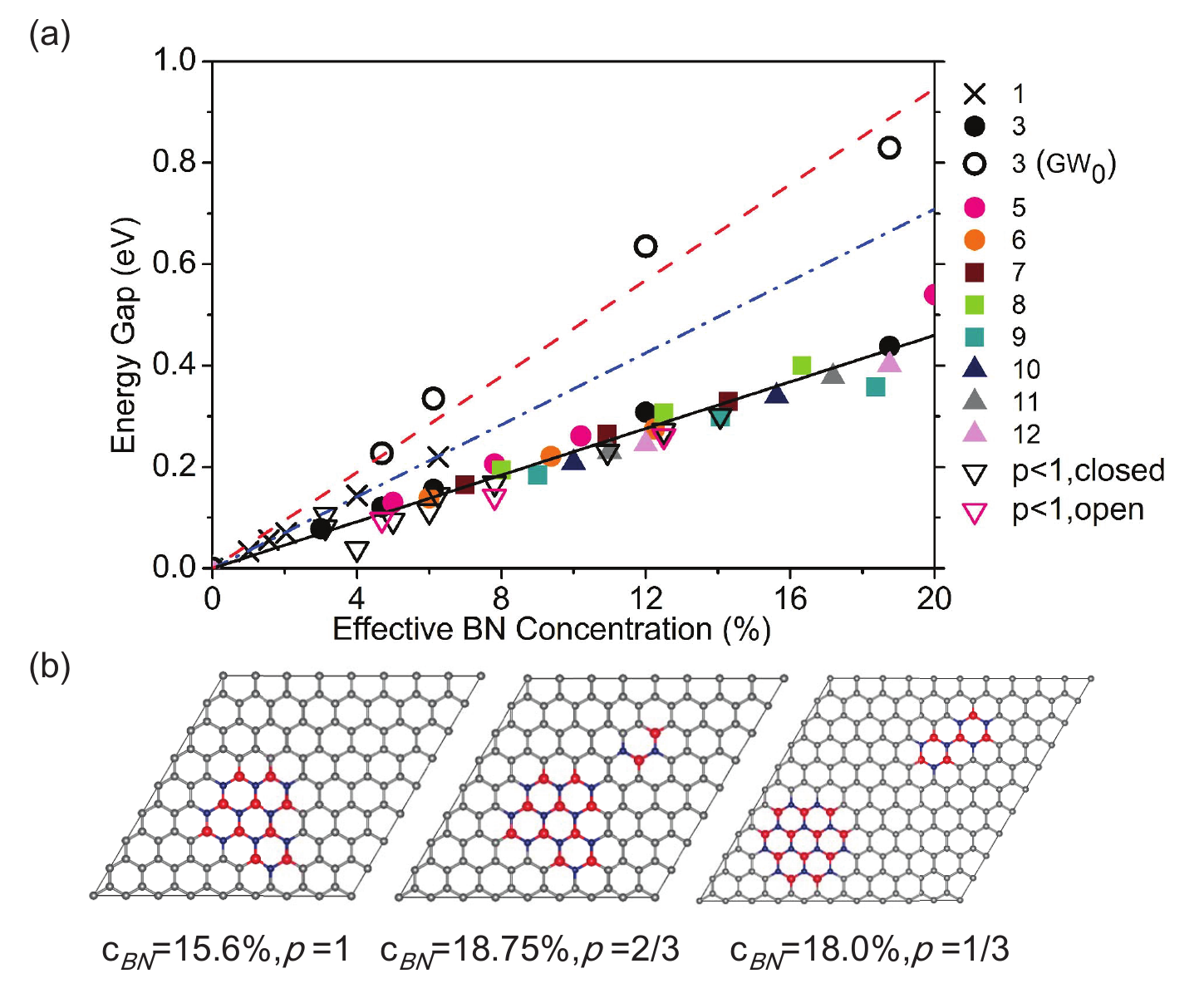}
\caption{Band gaps of $h$-BNC as a function of effective BN doping concentration. (a) Energy gaps obtained for lowest-energy, optimized BN structures as shown in Figure 1. 
The effective BN concentration and the sublattice polarization $p$ are defined in the text. The DFT energy gaps for coherent configurations ($p=1$) with lowest-energy, closed BN structures are marked by solid symbols, and fitted by a black solid line with a slope of 2.3 eV. The numbers shown in the legends indicate the number of B-N pairs. The DFT energy gaps of a single, dispersed B-N pair are marked by crosses and fitted by a blue dash-dotted line with a slope of 3.5 eV. The GW quasiparticle energy gaps for one BN ring ($\text{B}_3\text{N}_3$) in 4$\times$4, 5$\times$5, 7$\times$7, and 8$\times$8 supercells are marked by open circles and fitted by a red dashed line with a slope of 4.7 eV. The empty inverted black and red triangles represent energy gaps for non-coherent configurations ($p<1$) with closed and open BN structures, respectively.
(b) Examples of coherent ($p$ = 1) and non-coherent ($p$ $<$ 1) doping structures, labeled by the total BN concentration and sublattice polarization $p$.}
\label{dftgap}
\end{figure*}

\subsection{Energy Gaps}
Next we calculate the energy gap within density functional theory (DFT) for the lowest-energy BN structures in Fig. 1 using different supercells (4$\times$4, 5$\times$5, 7$\times$7, 8$\times$8, and 10$\times$10) corresponding to different BN concentrations up to 20$\%$. 
We will show in the next section that the unfolded energy bands in the $h$-BNC systems do mimic those of pristine graphene.
The energy-gap results are shown by different solid colored symbols in Fig. 3(a). It is noticeable that the variation reveals a distinct linear dependence of the energy gap with respect to the B-N pair concentration $E_g\propto c_{BN}$. The linear fitting of these lowest-energy, closed BN structures (black solid line) gives a slope of 2.3 eV, which is 50$\%$ of the DFT band gap of $h$-BN ($4.6$ eV). The calculated energy gaps for dispersed single B-N pairs are shown by crosses in Fig. 3(a). They also exhibit a linear behavior, but the slope of the linear fit (blue dash-dotted line) is 3.5 eV.
Note that the physical effect of BN embedding in graphene is distinctly different from that of creating antidots, as will be discussed later. 

The insertion of BN patches keeps the honeycomb lattice intact, but inevitably creates an effective on-site energy difference between the two sublattices. The existence of an on-site energy difference between the two sublattices 1 and 2 of a honeycomb lattice can be modeled by a tight-binding Hamiltonian. With an on-site energy difference $\Delta=\varepsilon_1-\varepsilon_2$ and a nearest-neighbor hopping $t$, the energy eigenvalues are:
\begin{eqnarray}
\varepsilon_\textbf{q}^{\pm}=\pm\sqrt{t^2|\gamma_\textbf{q}|^2+\Delta^2/4}\;,
\end{eqnarray}
where $\gamma$ is the nearest-neighbor phase factor, and $\textbf{q}$ is a wave vector away from the Dirac points K or K$'$. This results in an energy gap of $E_g=\varepsilon^+_0-\varepsilon^-_0=\Delta$ at the Dirac point ($\gamma_{\textbf{q}=0}=0$), proportional to the on-site energy difference.
The fact that our calculated band gap for many different configurations in Fig. 3(a) is only proportional to the number of B-N pairs suggests that the dispersed local BN dopants effectively introduce an average on-site energy difference for the whole layer and that the strength of this average on-site energy difference is proportional to the concentration of BN dopants.

In the discussions above, we have considered configurations in Fig. 1 where all B atoms are on one type of the sublattices and all N atoms on the other, the so-called ``coherent'' configuration. It could happen that different regions of BN dopants occupy opposite sublattices in the samples, which is denoted as a ``non-coherent'' configuration. The compensation effect is expected to reduce the effective on-site energy difference between the two sublattices. We denote the number of B-N pairs with B atoms on sublattice 1 and N atoms on sublattice 2 by $N^{12}$ and similarly the number of B-N pairs with B atoms on sublattice 2 and N atoms on sublattice 1 by $N^{21}$. These two opposite sublattice arrangements for BN dopants coexist in a non-coherent configuration. Therefore, a sublattice polarization can then be defined as $p=|N^{12}-N^{21}|/(N^{12}+N^{21})$, with $p$ = 1 for a coherent configuration and $p$ $<$ 1 for a non-coherent configuration. The average on-site energy difference for the whole layer could be related to an effective BN concentration given by 
$\Tilde{c}_{BN}$ = $p\,c_{BN}$. We have tested this idea by including a few non-coherent configurations in the calculations with and without closed hexagonal rings. Some of the examples are shown in Fig. 3(b), and discussions of more general configurations will be given later. The calculated energy gaps for these non-coherent configurations are shown by open inverted triangles in Fig. 3(a). The calculated data points ($p$ $<$ 1) fall on the same linear curve determined by the band gaps of coherent configurations ($p$ = 1), indicating that the linear relation is robust and insensitive to local BN structures. The current finding of an induced effective on-site energy difference between the two sublattices that determines the overall band gap is expected to be applicable to dispersed BN patches in real samples because of the averaging effect. This particular finding lays the foundation for the tunable electronic and topological properties of $h$-BNC. In addition, the slope of the graphene linear bands is slightly modified in $h$-BNC as will be discussed in the next section.

Since it is well known that the DFT Kohn-Sham gap is smaller than the true quasiparticle gap, we have performed GW$_0$ calculations for the one-BN-ring structure ($\text{B}_3\text{N}_3$) in 4$\times$4, 5$\times$5, 7$\times$7, and 8$\times$8 supercells. 
The bandgap results are shown by open circles in Fig. 3(a) that also follow a linear curve. Its slope is found to be 4.7 eV, about 2 times larger than the DFT value of 2.3 eV. This gives extrapolated gap values of 100 and 300 meV for 2$\%$ and 6$\%$ BN doping, respectively. These values are in the same order of magnitude as those measured (200 and 600 meV for 2$\%$ and 6$\%$ BN, respectively) \cite{Chang13}, although the BN patches may not be homogeneously dispersed in their CVD samples.

\subsection{Unfolded Band Structure}

\begin{figure}[htb]
    \centering
    \includegraphics[width=3.5in]{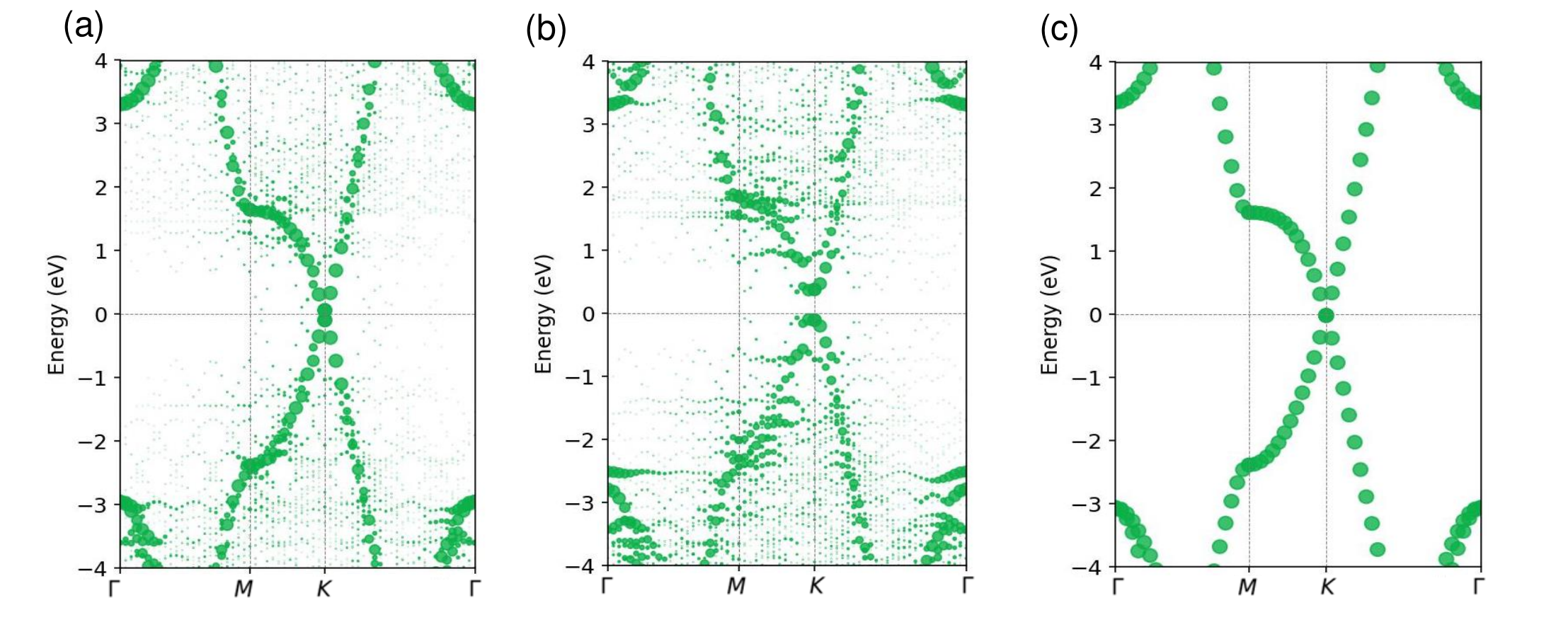}
    \caption{(a) Unfolded band structure of $h$-BNC for $\text{B}_3\text{N}_3$ in a 7$\times$7 supercell ($c_{BN}=6.1\%$). (b) Unfolded band structure of $h$-BNC for $\text{B}_8\text{N}_8$ in a 7$\times$7 supercell ($c_{BN}=16.3\%$). (c) Unfolded band structure of pristine graphene calculated using a 7$\times$7 supercell. }
    \label{fig:unfolded_band}
\end{figure}

Band unfolding calculations of $h$-BNC were performed using DFT with VASPKIT \cite{vaspkit} to study the overall impact of BN doping on the band structure of graphene. The unfolded band structures of $h$-BNC with 6.1\% BN concentration and 16.3 \% BN concentration are shown in the Fig.\;\ref{fig:unfolded_band}(a) and Fig.\;\ref{fig:unfolded_band}(b), respectively. The former $h$-BNC configuration contains a $\text{B}_3\text{N}_3$ in a 7$\times$7 supercell, while the latter contains a $\text{B}_8\text{N}_8$ in the same supercell. In comparison, the unfolded band structure of pristine graphene calculated using the same size of supercell is shown in Fig.\;\ref{fig:unfolded_band}(c).  

Our result shows that the BN dopant opens an energy gap while preserving the band dispersion away from the gap, indicating that BN dopant serves as a perturbation to the graphene band structure. At a low doping of 6.1\%, as shown in Fig.\;\ref{fig:unfolded_band}(a), an energy gap of about 0.15 eV is opened. On the other hand, the energy gap is about 0.4 eV in Fig.\;\ref{fig:unfolded_band}(b) with a large doping level of 16.3\%, and some additional BN defect states can be identified near the $\Gamma $ point, which are nearly 3 eV away from the Fermi level. Therefore, if we focus on the low BN concentration range of less than 10\%, it is expected that the characteristics of graphene bands can be preserved.

\subsection{Effect of BN Embedding}
\begin{figure}[ht]
  \centering
  \includegraphics[width=3.4in]{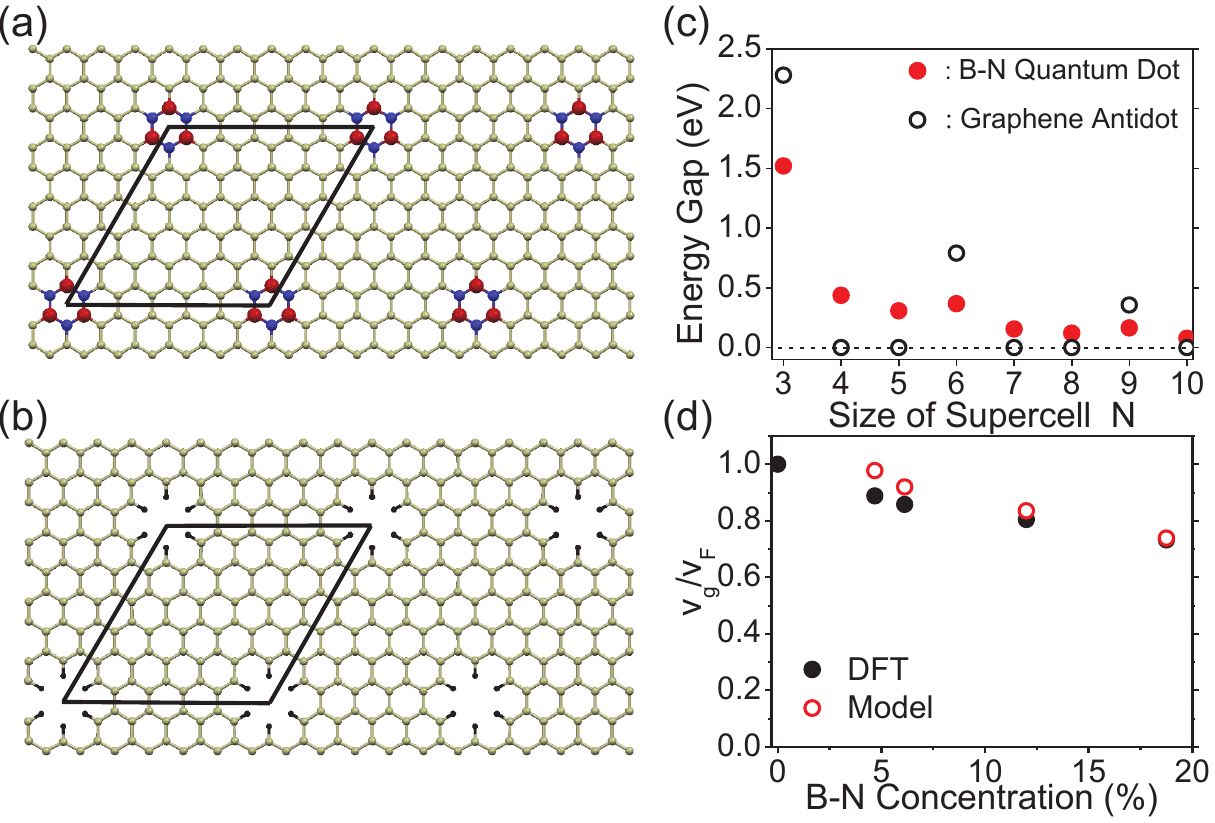}
  \caption{(a) Inserting a BN ring (creating a B-N quantum dot) and (b) removing C atoms (forming graphene antidots), as shown in the same 6$\times$6 graphene supercell. (c) Corresponding energy band gaps as a function of the size of the supercell.(d) Average group velocity of the graphene bands.} \label{qd}
\end{figure}

The embedding effect includes two aspects: the removal of C atoms and the insertion of BN patches. The former creates voids in graphene, and the later breaks the average sublattice symmetry. These two aspects introduce distinct effects in band gap opening as shown in Fig. \ref{qd}. We examine the energy gaps created by periodic antidots in graphene, in which one hexagonal ring of six C atoms per supercell is removed with the dangling bonds passivated by hydrogen. This opens an energy gap only for graphene antidots with a supercell size being a multiple of 3$\times$3, as shown Fig. \ref{qd}(c). Since the sublattice symmetry remains in the antidots lattice, it is the intervalley coupling, namely, the chiral symmetry breaking that lifts the degeneracy at Dirac points. In supercells of multiples of 3$\times$3, the band folding maps Dirac points K and K$'$ to $\Gamma$. The opening of the energy gap is determined by the periodic defect potential; a constructive (band gap opening) or destructive (band gap closing) interference is determined by the periodicity and symmetry of supercells and the defect structure factor at Dirac points \cite{Dvorak14,Zhang11,Lee11}. In contrast, an energy gap develops for the BN ring ($\text{B}_3\text{N}_3$) as shown in Fig. \ref{qd}(c). This indicates that the energy gap mainly arises from the local sublattice symmetry breaking, although a slightly bigger gap is found when the size of supercells is a multiple of 3$\times$3. This additional contribution is from chiral symmetry breaking, introducing an intervalley coupling between two valleys due to band folding. We have avoided using this group of supercells in our band gap calculations.

Since the slope of the linear bands (Fermi velocity) has a directional dependence reflecting the symmetry of the supercell \cite{Park08}, we therefore average over the group velocities determined at 0.2 eV above the conduction band edge along different directions around the original Dirac point. The results are plotted in Fig. \ref{qd}(d), showing that the renormalization is enhanced with increasing BN substitution. In the continuum limit where $|\textbf{q}|a\ll 1$, the group velocity obtained by $v_g=\hbar^{-1}\partial \varepsilon_\textbf{q}/\partial \textbf{q}$ will be
\begin{eqnarray}
v_g/v_F=\hbar v_F \textbf{q}/\sqrt{\hbar^2v_F^2q^2+\Delta^2/4}
\end{eqnarray}
where $a$ is the lattice constant, $v_F=3|t|a/2\hbar$ is the Fermi velocity of pristine graphene, and $\Delta$ is the average on-site energy difference. With $\Delta$ determined from the gap variation in Fig. \ref{dftgap}, the results from the continuum model agree well with DFT values, as shown in Fig. \ref{qd}(d). This again confirms that the embedded BN patches introduce an average on-site energy difference in the two sublattices of graphene.

\subsection{General Doping Configurations}
The artificial periodicity in a finite supercell used in previous sections may influence the physical results. Therefore, we construct a tight-binding (TB) model with parameters determined by fitting to DFT calculations in order to simulate the general doping configurations in larger supercells, such as 100 $\times$100. The TB model takes into consideration only the nearest-neighbor couplings, which are sufficient to capture the essential physical feature we are interested in (namely, the band gap opening).
The details of the fitting results between TB and DFT calculations are provided in Appendix and the TB parameters are given in Tabel \ref{table:1}.

\begin{figure}[ht]
  \centering
  \includegraphics[width=3.0in]{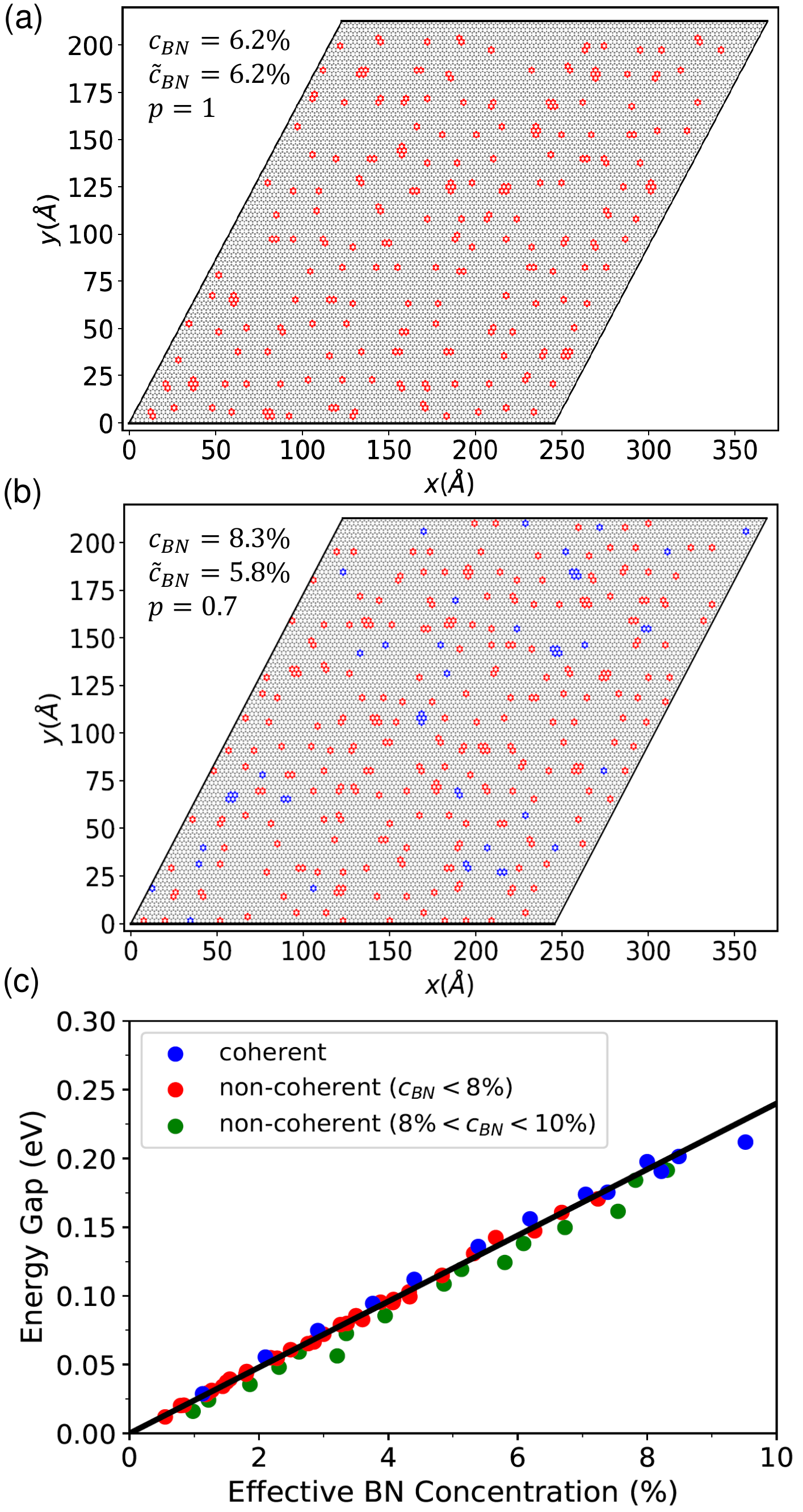}
  \caption{(a) An example of a coherent $h$-BNC configuration in a 100$\times$100 supercell. All of the BN dopants are marked in red, indicating that they have the same sublattice arrangement. (b) An example of a non-coherent $h$-BNC configuration in a 100$\times$100 supercell. The BN dopants with opposite sublattice arrangements are marked in red and blue, respectively, and they are dispersed in the sample.  (c) Calculated band gaps as a function of the effective BN concentration ($\Tilde{c}_{BN}$ = $p\,c_{BN}$) by the TB method for 100$\times$100 supercells. The sublattice polarization $p$ of non-coherent configurations ranges from 0.1 to 0.95. The data are fitted by a black solid line with a slope of 2.4 eV.} \label{dots}
\end{figure}

In our investigation of general $h$-BNC configurations, the total doping concentration is below 10\%. To best describe the $h$-BNC systems at low doping, we distribute the energetically optimized 3, 5, and 8 B-N pairs irregularly throughout the sample, since these closed structures have the lowest formation energy per B-N pair. In addition, the numbers of 3, 5, and 8 B-N pairs are chosen to satisfy $N_3>N_5>N_8$ because small BN patches are more likely to form at low doping. The examples of general coherent and non-coherent configurations are shown in Figs. 6(a) and (b), respectively. For non-coherent configurations, the BN dopants with opposite sublattice arrangements (marked in red and blue) are randomly chosen in positions, resulting in a dispersed distribution. 

We then calculate the energy gaps of various $h$-BNC coherent and non-coherent configurations as a function of the effective BN concentration ($\Tilde{c}_{BN}$ = $p\,c_{BN}$) using the TB model, and the results are shown in Fig. 6(c). For non-coherent configurations, the sublattice polarization $p$ considered covers a wide range of 0.1 to 0.95. All the data points are well fitted on the linear line for a total concentration below 8\%. As the total doping concentration raises between 8\% to 10\%, the energy gaps slightly drop below the linear line (green data points). In general, we conclude that the linear dependence between the energy gap and the effective BN concentration is preserved for $c_{BN}<10\%$.

The linear variation of the energy gap with the effective BN concentration highlights the importance of controlling the sublattice polarization of the BN dopants. Growing a  coherent $h$-BNC sample with $p=1$ is desirable as it maximizes the energy gap at a given total doping concentration. 
A recent experimental work reported the successful growth of wafer-scale single-crystal $h$-BN monolayers on Cu(111) \cite{Chen2020}. In the initial stage, the epitaxial growth is enhanced by lateral docking of $h$-BN to Cu (111) steps, generating mono-oriented $h$-BN triangular flakes on the substrate. This suggests the feasibility of growing oriented, low-concentration BN patches by controlling the precursor flow rate and duration. It can be followed by a comprehensive coverage of the remaining region with graphene. Thus a coherent $h$-BNC film could become achievable through this two-step growth process.


\subsection{Topological Properties}
\begin{figure}[ht]
  \centering
  \includegraphics[width=3.4in]{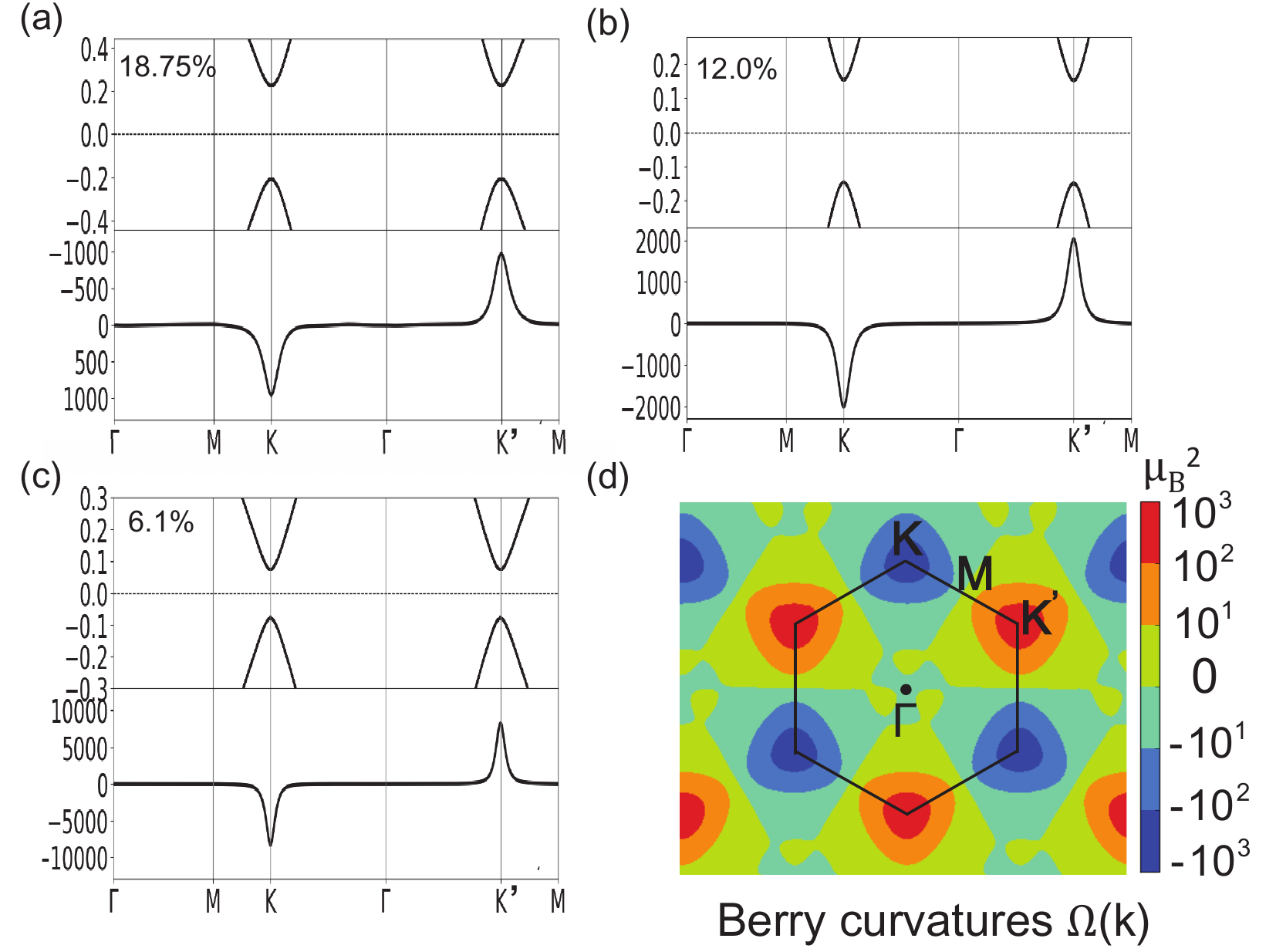}
  \caption{Topological properties of $h$-BNC. Berry curvatures (lower panel) and corresponding band structure (upper panel) by first-principles calculations in the vicinity of K and K$'$ valleys for the $h$-BNC monolayer with different BN concentrations: (a) 18.75$\%$, (b) 12$\%$, and (c) 6.2$\%$ evaluated with a ring-shaped $\text{B}_3\text{N}_3$ patch in 7$\times$7, 5$\times$5, and 4$\times$4 graphene supercells, respectively. (d) The Berry curvatures (-$\Omega_z$) of the $h$-BNC monolayer in a 2D k-plane for 18.75$\%$ BN. }\label{bncberry}
\end{figure}
\begin{figure}[ht]
\centering
\includegraphics[width=3.4in]{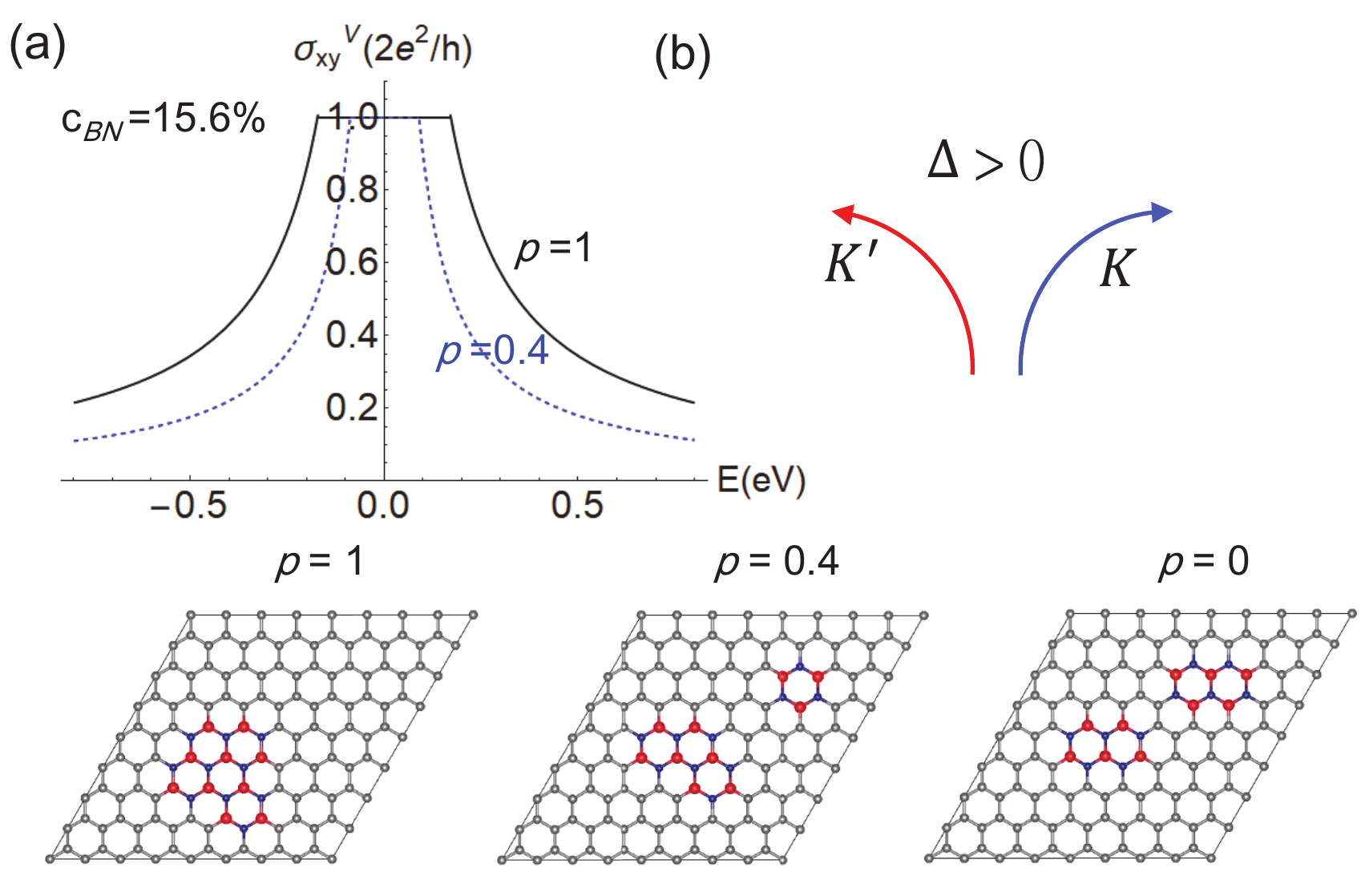}
\caption{Quantized valley Hall conductivity in the insulating region with a tunable plateau width. (a) Valley Hall conductivity of $h$-BNC calculated by the effective tight-binding model with a nearest-neighbor hopping of $t\simeq$2.6 eV and a = 2.46 $\textrm{\AA}$ for two different BN configurations with $c_{BN}$ = 15.6$\%$. The width of the insulating region is controlled by the BN concentration and sublattice polarization as determined by DFT calculations. (b) The valley current changes sign for opposite sublattice occupancy (as local effective onsite energy difference changes sign). The backscattering of single valley states cause the valley Hall conductivity to decrease outside the gap. The quantum valley Hall effect exists within the gap even for non-coherent sublattice substitution ($p<1$). The gap is closed at $p=0$.}\label{bncvhe}
\end{figure}
In the previous sections, we have shown that the electronic structure of the $h$-BNC monolayer can be described by the introduction of an average on-site energy difference on the two sublattices upon the BN insertion. Therefore, the substitution of carbon atoms in graphene by BN dopants breaks the inversion symmetry, giving rise to a nonzero Berry curvature, $\Omega(\vec{q})\neq\Omega(-\vec{q})$ and an anomalous velocity $-\dot{\vec{q}}\times\Omega(\vec{q})$. In Fig. 7, we show the Berry curvatures calculated from first principles by the construction of Wannier functions \cite{Mostofi08} for a few representative values of the BN concentration. The Berry curvatures are nonzero in all cases, exhibit peaks at the Dirac points (valleys), and gradually decrease as the BN doping level increases. 
The underlying physics can be described by an effective Hamiltonian of gapped graphene in the vicinity of Dirac points with crystal momentum $\vec{q}$ measured from the Dirac point: $H=\hbar v_F[\tau q_x\sigma_x+q_y\sigma_y+(\Delta/2)\sigma_z]$, where $\tau=\pm 1$ describes two valleys K and K$'$, $\sigma_i$ are Pauli matrices, and $\Delta$ is the on-site energy difference between two sublattices. The eigenvalues are $\varepsilon_q=\pm\Delta/2\sqrt{1+\lambda_D^2 q^2}$ with $\lambda_D=\hbar v_F/(\Delta/2)$.
The Berry curvature is valley dependent and is given by
\begin{eqnarray}
\Omega_\pm^\tau(\varepsilon_q)=\mp\tau\frac{(\hbar v_F)^2\Delta}{4\varepsilon_q^3}\;,
\end{eqnarray}
which contributes to the intrinsic Hall conductivity and changes sign between the K and K$'$ valleys \cite{Xiao99,Lensky15}. Our results from first-principles calculations are in excellent agreement with this simple Dirac model.
The intrinsic valley Hall conductivity (VHC) at $T = 0$ K is obtained from $\sigma_{xy}^V=\sum_s(\sigma_{xy}^{K,s}-\sigma_{xy}^{K',s})=(4e^2/\hbar)\int d^2q  \, \Omega(q)/(2\pi^2)$, with the spin index $s$ and the factor of 4 arising from the spin and valley degeneracy. The VHC includes contributions from the anomalous velocities of all occupied states below the Fermi energy. When the Fermi level is above the conduction band minimum $E_F\geq\Delta/2$, one has $\sigma_{xy}^V=(e^2/h)(\Delta/E_F)$ that decays as the energy moves away from the insulating regime. When the Fermi level is inside the gap $|E_F|\leq\Delta/2$, the VHC exhibits a quantized value, $\sigma_{xy}^V=2e^2/h$ \cite{Xiao99,Lensky15,Yamamoto15,Ando15}. It is shown previously that the plateau of valley Hall conductivity of a gapped graphene can survive from the short-range and long-range disorders in the absence of intervalley scattering \cite{Yamamoto15,Ando15,Cresti16}. With a tunable band gap and Berry curvature through the BN concentration and sublattice polarization in $h$-BNC, this system manifests itself as an ideal platform for realization of valleytronics and QVHE in single layer graphene, as shown in Fig. 8.

\section{Conclusions}
In conclusion, after determining the energetically optimized structures of BN patches embedded in graphene using first-principles calculations, we have systematically examined the mechanism of band-gap opening at low BN concentration. Furthermore, a tight-binding model is constructed to simulate general doping configurations in large supercells. We find a band gap variation that scales linearly with the effective BN doping concentration taking into account the sublattice occupation order, with a slope of about 4.7 eV based on the GW results. This indicates the existence of an effective and tunable site-energy difference between the two carbon sublattices in the whole system, which breaks the sublattice symmetry as well as the inversion symmetry, giving rise to nonzero Berry curvatures at Dirac points with a sizable band gap. This BN-embedded graphene system is a potential platform to explore topological valley transport properties and to provide a promising system for future valleytronics applications.

\begin{acknowledgements}
This work is supported by a Thematic Project at Academia Sinica (AS-TP-106-M07).
\end{acknowledgements}

\appendix*
\section{Tight-binding calculations}

\begin{figure}[ht]
\centering
\includegraphics[width=2.8in]{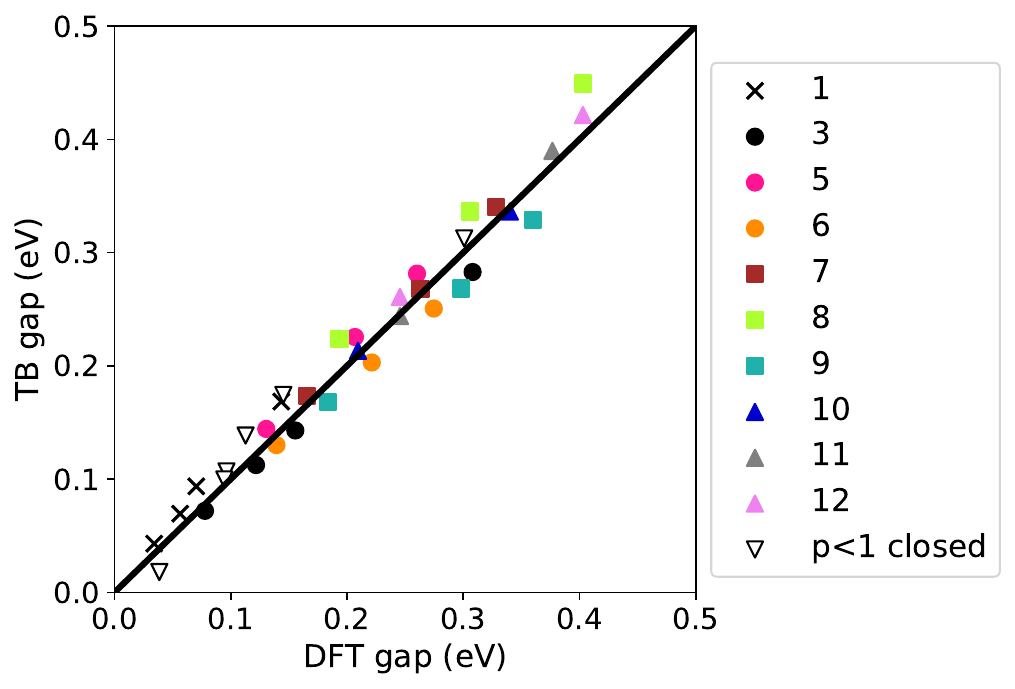}
\caption{Comparison of energy gaps from tight-binding (TB) calculations with those from DFT calculations for various $h$-BNC configurations used in the fitting. The supercell sizes are chosen to be 4$\times$4, 5$\times$5, 7$\times$7, 8$\times$8, and 10$\times$10. The numbers in the legend refer to the numbers of B-N pairs for energetically optimized structures. The solid data points are coherent configurations ($p=1$), while the empty inverted black triangles indicate the non-coherent configurations ($p<1$) with closed BN structures.}
\label{fig:gap_fit}
\end{figure}

\begin{figure}[h!]
    \centering
    \includegraphics[width=2in]{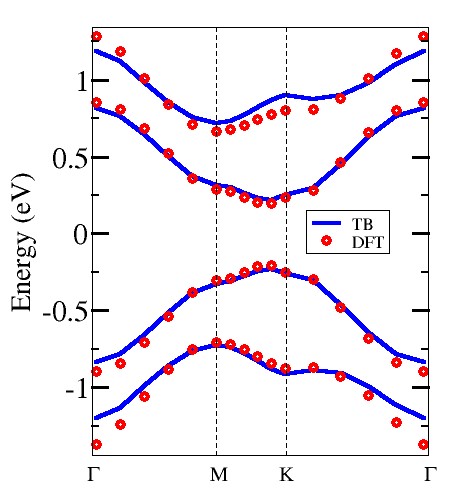}
    \caption{Low-energy band structure for optimized 8 B-N pairs in a 7$\times$7 supercell. Results from the TB model agree with those from DFT calculations.}
    \label{fig:band}
\end{figure}

In Fig.\;\ref{fig:gap_fit}, the energy gaps from tight-binding (TB) calculations are compared with those from DFT calculations for various $h$-BNC configurations used in the fitting, and a good agreement can be found.
We also calculated the $h$-BNC band structure using the TB model and found that it agrees well with the DFT result. An example is shown in Fig.\;\ref{fig:band}. With the nearest neighbor couplings, the TB model is sufficient to capture the essential electronic properties of $h$-BNC.  

\begin{figure}[htb]
    \centering    
    \includegraphics[width=3in]{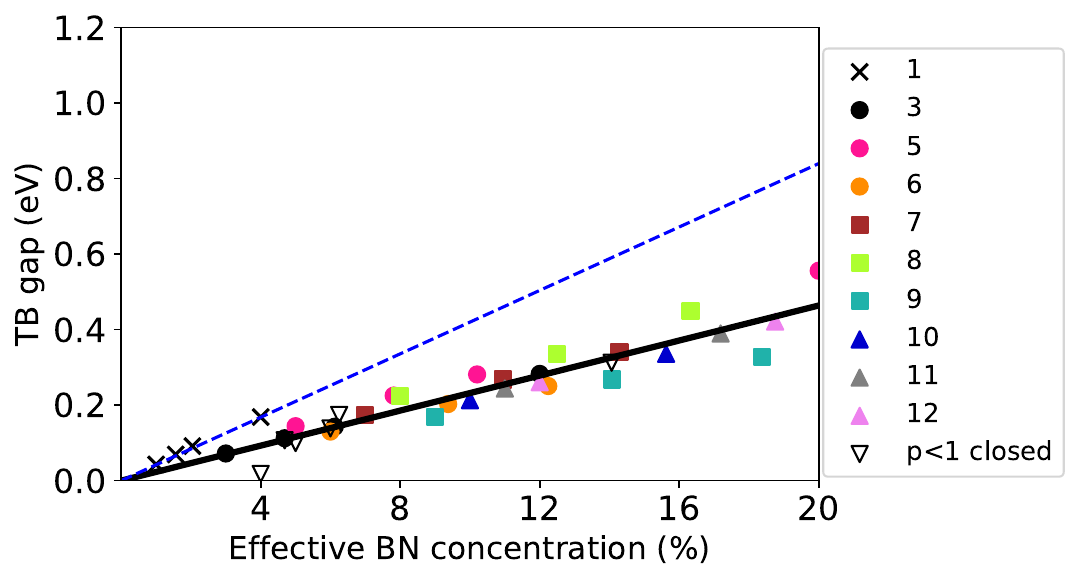}
    \caption{Band gaps of $h$-BNC as a function of effective BN concentration calculated by the tight-binding model. The supercell sizes are chosen to be 4$\times$4, 5$\times$5, 7$\times$7, 8$\times$8, and 10$\times$10. The numbers in the legend refer to the number of B-N pairs. The solid data points are coherent configurations ($p=1$), while the empty inverted black triangles indicate the non-coherent configurations ($p<1$) with closed BN structures. These data are fitted by a black solid line with a slope of 2.3 eV, which agrees well with the conclusion from DFT calculations in Fig.~3(a). On the other hand, energy gaps for configurations with a single B-N pair are marked by black crosses and follow the blue dashed line with a slope of 4.1 eV.}
    \label{fig:TB_gaps}
\end{figure}

In small supercells, the TB calculations give a linear dependence of the energy gaps with respect to the effective BN doping concentration for various configurations as shown in Fig.\;\ref{fig:TB_gaps}. 
The data are fitted by a black solid line with a slope of about 2.3 eV, which agrees with the conclusion from DFT calculations in Fig.~3(a). On the other hand, the gaps of configurations with a single B-N pair follow a linear line with a slope of 4.1 eV, which is larger than the value of 3.5 eV in DFT calculations. It suggests that the TB parameters for single B-N pairs may be different from other closed BN structures. In this study, we neglect the single B-N pair configurations, since they may be less likely to form in experiment because the energy per B-N pair is much higher than that of other closed structures.


\bibliography{bnc_chuu}

\end{document}